# A Centroiding Algorithm for Point-source Trails

Linpeng Wu[1], Qingfeng Zhang[1,2], Valéry Lainey[3], Nick Cooper[4], Nicolas Rambaux[3], and Weiheng Zhu[1,2]
[1] Department of Computer Science, Jinan University, Guangzhou 510632, People's Republic of China; tqfz@jnu.edu.cn
[2] Sino-French Joint Laboratory for Astrometry, Dynamics and Space Science, Jinan University, Guangzhou 510632, People's Republic of China
[3] IMCCE, Observatoire de Paris, PSL Research University, CNRS UMR 8028, Sorbonne Université, UPMC, Univ. Lille 1, 77 avenue Denfert-Rochereau, Paris 75014, France
[4] Department of Physics & Astronomy, Queen Mary University of London, Mile End Rd, London, E1 4NS, UK

## Abstract

Astrometric measurements are significantly challenged by the relative motion between the point source and the telescope, primarily due to the difficulty in accurately determining the position of the point source at the mid-exposure moment. Especially when the trail is irregular in shape or results from nonuniform relative motion, determining the centroid of such a trail becomes significantly more challenging. To address this issue, a new centroiding algorithm for point-source trails has been developed. This algorithm employs a piecewise linear model to approximate the irregular trajectory of a point source. An estimated intensity distribution of the trail is constructed by integrating the point-spread function with the approximated trajectory. The cost function is defined as the difference between the estimated and observed trail intensity distributions, with an added smoothness constraint term. Optimizing this cost function yields a refined trajectory fit. A coarse-to-fine iterative approach is used to progressively converge on the true trajectory of the point source, ultimately determining both the trail's centroid and the trajectory of the point source. The efficacy of the algorithm is validated using synthetic images. Furthermore, this technique is applied to Cassini Imaging Science Subsystem images of several inner Saturnian satellites, successfully processing 267 astrometric observations. The results demonstrate the effectiveness of the algorithm in real astronomical applications.

*Unified Astronomy Thesaurus concepts:* Observational astronomy (1145); Astronomy image processing (2306); Astrometry (80); Ephemerides (464); Saturnian satellites (1427)

Materials only available in the online version of record: machine-readable table

## 1. Introduction

Trailing imaging of observed objects is a common occurrence when there is relative motion between the objects and the telescope. For instance, streaks often appear when artificial satellites, space debris, or near-Earth objects are observed from the ground, because of their rapid motion relative to the Earth. Similarly, during space-based observations, trails may appear in the image due to either camera instability or errors in tracking the target. In Cassini Imaging Science Subsystem (ISS) images, the trailing effect is prominently observed on background reference stars and/or the moving targets during long exposures (tens of seconds) when the camera tracks a moving target. Figure 1 illustrates examples of such trailing, where the trajectories are irregular in shape and the relative motion may be nonuniform. The irregular shape and nonuniform motion in trailing increase the difficulty of accurately determining the centroid of the trail, which refers to the position of the trailed source at the mid-exposure moment. Regardless of the nature of the trailing, accurately determining the centroids of trails is a fundamental task in astrometry. The motivation for our research is to develop a centroiding algorithm capable of effectively handling both regular and irregular trailing patterns.

Various approaches have been used to detect trails and determine their centroids in charge-coupled device (CCD) images. M. Levesque (2009) employed iterative matching filtering with line segment templates for streak localization after image preprocessing. However, obtaining an accurate center was not involved. P. Vereš et al. (2012) constructed a linear trail model that incorporates a Gaussian point-spread function (PSF) and a consistent asteroid motion rate and fitted the model to real asteroid images from the Pan-STARRS1 survey to obtain the centers of streaks. R.-Y. Sun et al. (2013) utilized morphological techniques for linear streak detection and applied the modified moment method to compute their centers. J. Virtanen et al. (2016) developed the StreakDet pipeline, enabling streak detection and astrometric reduction in low-signal-to-noise-ratio (SNR) scenarios. Their approach involved segmentation, streak classification, and the use of a 2D PSF along a line or arc to estimate streak parameters, including the center position. W. Fraser et al. (2016) introduced the TRIPPy Python package for trailed image photometry, leveraging the "pill aperture" technique for flux measurements. TRIPPy focused on photometry instead of centroiding. B. Sease et al. (2017) suggested a method for identifying the endpoints of streaks through corner detection and subsequently calculated the center positions using these endpoints. G. Nir et al. (2018) proposed an optimal approach by cross-correlating images with a line template that is expanded by the system's PSF to detect streaks, getting the position of one streak by using the Radon transform. G. Privett et al. (2019) gave a method for streak centroiding in wide-angle CCD images. It utilized a Hough transform technique to detect the presence of streaks. Additionally, a Gaussian PSF model along a line was employed to determine the center of each streak. A. Vananti et al. (2020) presented a spatial filter designed to detect faint, streak-like trails in images, working under the assumption of linear streaks. Their work tackled the challenge of identifying streaks that have different orientations and lengths within a single image. R. Haussmann et al. (2021) introduced a passive optical system for detecting space debris. In this system, streaks were identified by the shape parameters of regions of interest (ROIs) detected after image enhancement and binarization. Intensity moments were then used to determine the central positions of the streaks. B. Lin et al. (2022) employed a three-domain clustering method for streak detection and utilized linear fitting to estimate streak parameters, without focusing on the centroiding of streaks and under the assumption that streaks are linear. J. Du et al. (2022) offered a template-matching approach to detecting trailed sources, where the template matching could identify both linear and distorted streaks. A modified moment method was employed to compute the centroids of the streaks. This centroiding approach was suitable for linear streaks or those with slight distortion. However, it could not accurately determine the centroids for streaks with mild or severe distortions. Q.-F. Zhang et al. (2023) proposed one image-processing method to detect the linear streaks in Cassini ISS images and used the modified moment method to obtain their centers.

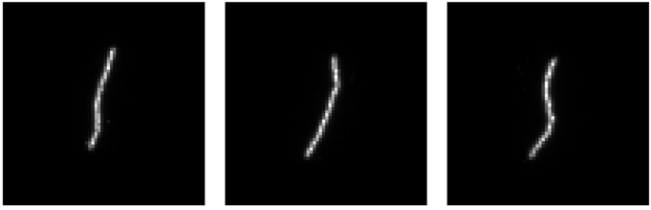

**Figure 1.** Trailed reference stars in Cassini ISS images, showing irregular and varying trajectories due to unstable camera movement.

All the aforementioned techniques represent traditional approaches that primarily rely on traditional digital image-processing techniques for trail detection, employing methodologies such as the Hough transform, Radon transform, template convolution, and corner detection. Some studies focus solely on streak detection, while others extend to obtaining accurate streak centroids post-detection, utilizing centroiding methods including modified moment techniques, PSF fitting, and endpoint localization (e.g., using the midpoint between two endpoints as the streak's centroid). However, these techniques tend to be scenario-specific in their applications. Furthermore, the majority of research efforts have focused on handling linear streaks. Only the methods proposed by J. Virtanen et al. (2016) and J. Du et al. (2022) can obtain the centroids of slightly curved streaks, but accurately determining the centroids for irregular trails remains a challenge.

Recently, deep learning methods have been applied to streak detection (D. A. Duev et al. 2019; L. Varela et al. 2019; P. Jia et al. 2020; J. Xi et al. 2020; R. Haussmann et al. 2021; A. A. Elhakiem et al. 2023). These studies primarily focus on identifying regions containing streaks rather than precisely determining their centroids. Additionally, these neural networks are limited to detecting linear streaks. Although deep learning techniques are promising for streak detection, they present certain limitations compared to traditional detection methods. For example, deep learning methods require extensive labeled data for training neural networks. Moreover, current neural networks designed for streak detection lack generalization capabilities and are often tailored to specific scenarios. Importantly, they are unable to accurately estimate the centers of streaks.

In this paper, an algorithm is proposed to determine the centroids of point-source trails, applicable to both linear and irregularly curved trails. The performance of this algorithm is first tested through experiments with synthetic images. It is then applied to the astrometry of Cassini ISS images containing trailed stars of several major Saturnian satellites. The remainder of this paper is organized as follows. Section 2 introduces the principle of our algorithm and its implementation details. Section 3 describes the experiments for testing the performance of our centroiding algorithm by using synthetic images. Section 4 presents the application to the astrometry of Cassini ISS images. A discussion is given in Section 5. The conclusions are drawn in Section 6.

## 2. Method

To accurately determine the centroid of a trail, regardless of whether the trail is linear or irregularly curved, we employ a multistep approach. This involves: (1) approximating the trail's trajectory using a piecewise linear model; (2) constructing an estimated intensity distribution by integrating the PSF with the approximated trajectory; (3) optimizing a cost function to refine the trajectory fit; and (4) employing a coarse-to-fine strategy to iteratively converge on the true trajectory. Finally, the centroid of the trail is determined based on the refined trajectory. In the following, we elucidate the principles and implementation of this method.

### 2.1. Principle of Centroiding of Point-source Trail

#### 2.1.1. PSF of Point Source

The PSF is conventionally defined as the spatial response of the detector resulting from a point source of light traversing through the optical system. For a stationary point source captured in a CCD image, unaffected by any relative motion to the camera, its intensity profile can be characterized by the following equation (J. Anderson & I. R. King 2000):

$$I(\boldsymbol{i}) = f_\star \cdot \psi(\boldsymbol{i} - \boldsymbol{X}_\star) + b(\boldsymbol{i}). \qquad (1)$$

Here, $\boldsymbol{i}$ represents the coordinate of a pixel, $I$ is the intensity at that location, $\boldsymbol{X}_\star$ denotes the coordinates of the center position of the point source (e.g., star or asteroid), $f_\star$ stands for the flux factor, expressing the brightness of the point source, $b(\boldsymbol{i})$ is the background value at $\boldsymbol{i}$, and $\psi$ denotes the PSF, where the integration over the pixel's spatial extent is omitted for computational simplicity. In space-based astronomical image processing, it is common to use more refined subpixel-grid PSF models to approximate the continuous PSF, as detailed by J. Anderson & I. R. King (2000).

#### 2.1.2. PSF of Trail

When there is relative motion between an observed object and the camera, a point source will appear as a trail in the image. This trail can be either linear or curved. The motion of the point source projected onto the image plane during the exposure period is described by the trajectory $s(t)$, which also represents the movement of $\boldsymbol{X}_\star$. The exposure duration is denoted as $[-T, T]$. The centroid of the trail, $s(0)$, corresponds to the position of the point source at the midpoint of the exposure. However, $s(0)$ may not coincide with the geometric center of the trail if the point source's velocity varies during the exposure

The intensity distribution of the trail can be expressed by the following equation:

$$I(\boldsymbol{i}) = f_\star \int_{-T}^{T} \psi(\boldsymbol{i} - \boldsymbol{s}(t)) dt + b(\boldsymbol{i}). \quad (2)$$

To simplify the notation, we define

$$\Psi(\boldsymbol{i}, \boldsymbol{s}) = \int_{-T}^{T} \psi(\boldsymbol{i} - \boldsymbol{s}(t)) dt, \quad (3)$$

where $\Psi$ characterizes the intensity profile of a trail caused by a moving point source, referred to as the trailed PSF (tPSF). This function describes how the image of a point source is smeared due to its motion throughout the exposure period. Notably, $\Psi(\boldsymbol{i}, \boldsymbol{s})$ is functionally dependent on $(\boldsymbol{i} - \boldsymbol{s}(0))$. This means that if multiple trails share the same motion trajectory, differing only by translation, their intensity profiles will be identical. Here, the notation $\Psi(\boldsymbol{i}, \boldsymbol{s})$ is used for convenience and generality.

Thus, the intensity profile of the trail can be rewritten as

$$I(\boldsymbol{i}) = f_\star \cdot \Psi(\boldsymbol{i}, \boldsymbol{s}) + b(\boldsymbol{i}). \quad (4)$$

From the equation, we can conclude that the tPSF is crucial for computing the intensity distribution of a trail.

### 2.1.3. Cost Function

The residual intensity $\chi$ between the estimation and true value is described by the following equation:

$$\chi(\boldsymbol{i}, \hat{\boldsymbol{s}}) = f_\star \cdot \Psi(\boldsymbol{i}, \hat{\boldsymbol{s}}) + b(\boldsymbol{i}) - I_\mathrm{m}(\boldsymbol{i}), \quad (5)$$

where $I_\mathrm{m}(\boldsymbol{i})$ is the measured intensity of the pixel at $\boldsymbol{i}$ and $\hat{\boldsymbol{s}}$ represents an estimation of true trajectory $\boldsymbol{s}$.

Let $A$ represent the ROI, which includes all valid pixels related to the trailed source. Specifically, $A$ is the collection of pixels within which the trailed source's flux is integrated and measured. In the ROI $A$, we define a cost function, denoted as $L$, which quantifies the discrepancy between the estimated and actual intensity profile of the trailed source as follows:

$$L(\hat{\boldsymbol{s}}) = \sum_{\boldsymbol{i} \in A} (\chi(\boldsymbol{i}, \hat{\boldsymbol{s}}))^2. \quad (6)$$

This function serves as a metric for assessing the discrepancy between the estimated and the true trajectory. Typically, optimizing the cost function $L$ is expected to produce an estimated trajectory $\hat{\boldsymbol{s}}$ that approximates the true trajectory and aligns more accurately with the observed data. However, optimizing only $L$ may yield an unsuitable approximation that exhibits erratic or oscillatory behavior, deviating significantly from the expected smooth motion of the trailed source. To address this issue and impose a smoothness constraint on the estimated trajectory, we introduce an additional regularization term to the cost function, as follows:

$$L(\hat{\boldsymbol{s}}) = \sum_{\boldsymbol{i} \in A} (\chi(\boldsymbol{i}, \hat{\boldsymbol{s}}))^2 + |A| f_\star^2 \cdot (\lambda_n J_n(\hat{\boldsymbol{s}}) + \lambda_t J_t(\hat{\boldsymbol{s}})), \quad (7)$$

where $|A|$ is the number of available pixels in the ROI, $\lambda_n$ and $\lambda_t$ are the weighting coefficients, and $J_n$ and $J_t$ are energy terms in the normal and tangential directions of trajectory. They are defined as

$$\begin{cases} J_n(\hat{\boldsymbol{s}}) = \int_{-T}^{T} \|\hat{\boldsymbol{s}}''(t) \cdot \boldsymbol{e}_n(t)\|^2 dt \\ J_t(\hat{\boldsymbol{s}}) = \int_{-T}^{T} \|\hat{\boldsymbol{s}}''(t) \cdot \boldsymbol{e}_t(t)\|^2 dt \end{cases}, \quad (8)$$

where $\boldsymbol{e}_n(t)$, $\boldsymbol{e}_t(t)$ are unit vectors in the normal and tangential directions of the trajectory, respectively. $\hat{\boldsymbol{s}}''(t)$ is the acceleration of the estimated trajectory, which is detailed in the Appendix.

In Equation (7), the second term, known as the regularization term, combines the energy terms along the normal and tangential directions of the trajectory. Assuming the motion blur exhibits smoothness, this regularization component is designed to penalize trajectories demonstrating excessive curvature or rapid directional changes. By minimizing the overall cost function $L$, an estimated trajectory approximating the true path can be obtained, from which the centroid of the trail is derived. The computation of Equation (7) is detailed in the Appendix.

### 2.2. Piecewise Linear Model

As demonstrated by Equation (7), optimization methods can be utilized to achieve an estimate of the trajectory $\boldsymbol{s}(t)$. However, a crucial consideration is how to appropriately represent $\boldsymbol{s}(t)$. It is important to note that our objective is to determine its centroid, not its geometric center. Therefore, the expression for $\boldsymbol{s}(t)$ should incorporate components related to velocity. Following the approach suggested by S. Oh & G. Kim (2014), we use a piecewise linear function to represent the possibly nonuniform trajectory. Consequently, an estimated trajectory can be described as

$$\begin{cases} \hat{\boldsymbol{s}}(t) = \sum_{j=0}^{Q-1} g_j(t) \left( \boldsymbol{x}_j + \frac{t - t_j}{2T/Q} (\boldsymbol{x}_{j+1} - \boldsymbol{x}_j) \right) & t \in [-T, T] \\ g_j(t) = \begin{cases} 1, & t_j \leqslant t < t_{j+1} \\ 0, & \text{otherwise} \end{cases} \\ t_j = -T + \frac{2T}{Q} \cdot j & j = 0, 1, \dots, Q \end{cases}, \quad (9)$$

where $Q$ represents the total number of segments, which should be an even number for convenience, and $g_j(t)$ represents a boxcar function. The exposure duration is defined over the interval $[-T, T]$. The sequence of times $\{t_j\}$ is uniformly distributed within this interval. The position of the trailed source at time $t_j$ is denoted as $\boldsymbol{x}_j$, referred to as control points.

From Equation (9), it is evident that control points are crucial. As the time $t$ varies, the control points $\{\boldsymbol{x}_j\}$ determine the general shape of the trajectory. If the control points are sufficiently dense and accurate, the piecewise linear segments $\hat{\boldsymbol{s}}(t)$ will closely approximate the true trajectory. The control point $\boldsymbol{x}_{Q/2}$, corresponding to the time $t_{Q/2}$, will align precisely with the centroid of the trail $\boldsymbol{s}(0)$.

### 2.3. Coarse-to-fine Method

The key to accurately representing the trajectory with piecewise linear segments lies in the control points. If the control points are not precise enough or insufficient in number, merely optimizing the cost function will not enable these segments to approximate the true trajectory. Typically, even after optimizing the initial control points according to Equation (7), the piecewise line segments constructed by them still deviate significantly from the true path. To achieve the goal of closely approximating the true trajectory, we propose a coarse-to-fine approach. This method gradually refines the sequence of control points $\{x_j\}$ to enable the line segments to approximate the true trajectory.

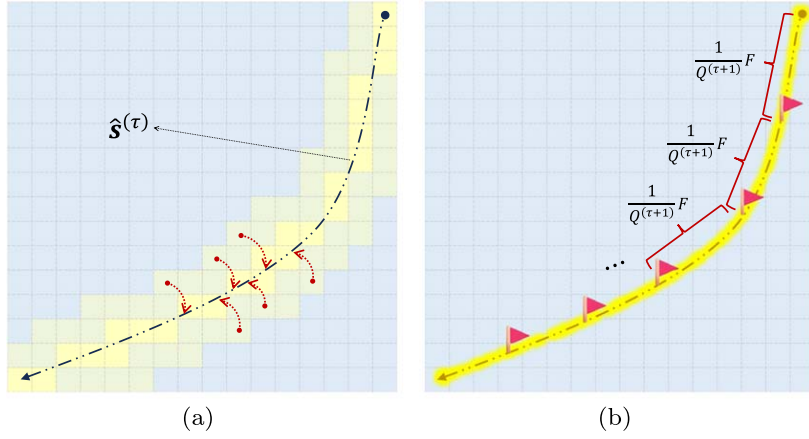

**Figure 2.** Illustration of control point refinement. (a) Construction of the curve-guided intensity distribution $H$. The dashed–dotted curve represents the trajectory $\hat{s}^{(\tau)}$. Each pixel in the ROI is scanned, and its intensity $(I_m(i) - b(i))$ is added to the nearest subpixel position on $\hat{s}^{(\tau)}$, as illustrated by the short arrows. $H$ is thereby created. (b) Equalization refinement process. The intensity distribution $H$ is divided into segments with equal cumulative intensity, $F/(Q^{(\tau+1)})$. The refined control points are positioned at the locations segmenting $H$, as illustrated by the flag icons.

This method comprises three main steps: (1) determining the number of control points for the next iteration; (2) constructing a curve-guided intensity distribution along the estimated trajectory; and (3) subdividing the curve-guided intensity distribution into segments by the number of control points to obtain the refined control points. The procedure is as follows:

1. Determining the number of the control points in next iteration. Let $(Q^{(\tau)} + 1)$ denote the number of control points in the $\tau$ iteration. We determine $(Q^{(\tau+1)} + 1)$ for the next iteration to be an odd number smaller than $(2Q^{(\tau)})$. This generally ensures that the control points are roughly doubled in density, while also allowing some deviation from the previous control points. The specific strategy for increasing the number of control points can vary, but our approach has been proven effective in practice.

2. Constructing the curve-guided intensity distribution $H$ along the estimated trajectory $\hat{s}^{(\tau)}$. During this process, each pixel in the ROI is scanned, and its intensity $(I_m(i) - b(i))$ is mapped to the nearest subpixel position on $\hat{s}^{(\tau)}$. If multiple pixels map to the same position on the curve, their intensities are summed. If a pixel is equidistant from several nearest positions on the curve, its intensity is evenly distributed among them. This redistribution continues for all pixels in the ROI, resulting in the intensity distribution $H$ along the curve. Figure 2(a) shows the process.

3. Obtaining the refined control points. The intensity distribution $H$ is divided into segments with equal cumulative intensity. For $H$, we first calculate its total intensity sum $F$, then determine the desired intensity sum $F_{\text{interval}} = F/(Q^{(\tau+1)})$ between each adjacent pair of control points. Starting from the first control point along the trajectory $\hat{s}^{(\tau)}$, the cumulative sum of $H$ is calculated. If the cumulative sum equals or approximates the desired intensity sum $F_{\text{interval}}$ for a segment, the corresponding position is identified as the next control point. This process is repeated until all control points along $\hat{s}^{(\tau)}$ have been obtained. Figure 2(b) illustrates the equalization refinement process.

### 2.4. Implementation

For a trail in the image, we employ the following seven steps to approximate its true trajectory and obtain the centroid of the trail:

1. Preprocess the image, determine the ROI of the trail, and compute its background intensity.
2. Provide an initial set of control points. We have designed a user interface that allows the user to manually specify a few initial control points.
3. Compute the tPSF along the trajectory $\hat{s}(t)$ described by the control points by using the known PSF, then calculate $f_\star$ based on the tPSF.
4. Construct the cost function $L(\hat{s})$ according to Equation (7) and minimize it to obtain the optimized control points.
5. Apply the coarse-to-fine method stated above to refine the control points.
6. Repeat steps 3–5 until the absolute difference between consecutive estimates of $\hat{s}(0)$ is less than a user-specified tolerance, typically set to 0.01. The iterative process also terminates if the number of control points surpasses $2N$ (where $N = 256$ in our implementation).
7. With the refined control points, compute the trajectory $s(t)$ and its centroid $s(0)$ using the piecewise linear approximation described earlier.

This iterative approach, incorporating user input, coarse-to-fine refinement, and cost function minimization, enables the estimation of the true trajectory and centroid of the trail, even in cases of irregular or complex motion patterns. In all the steps, some details should be noted.

In step 1, the rectangular region containing the trail is found by the common image-processing techniques described in Q.-F. Zhang et al. (2023). The background intensity $B$ is determined as the average of the sigma-clipped values within the rectangle. A mask defining the trail's shape is generated by region segmentation and morphological operation to mark the pixels belonging to the ROI $A$. In general, all pixels within the ROI share the same background intensity.

In step 2, the initial control points are provided. Given the irregularity and variability of trails, a graphical user interface (GUI) has been developed to facilitate this process. For simple trails, typically three control points (a start point, midpoint, and endpoint) are sufficient. For more complex trails, the GUI allows users to specify additional control points as needed. The GUI also enables the selection of clean and bright trails, discarding those that are interfered with by other objects or severely degraded by noise.

In step 3, tPSF is computed by numerical integration, as follows:

$$\Psi(\boldsymbol{i}, \hat{\boldsymbol{s}}) = \frac{T}{N} \sum_{n=-N}^{N-1} \psi\left(\boldsymbol{i} - \hat{\boldsymbol{s}}\left(\frac{T}{N}\left(n + \frac{1}{2}\right)\right)\right). \quad (10)$$

Theoretically, increasing $N$ enhances the precision of numerical integration but proportionally raises the computational cost. In our analysis of trails in Cassini ISS images, we empirically set $2N = 512$. The algorithm will terminate if the number of control points exceeds $2N$.

The $f_\star$ in the ROI is estimated by

$$f_\star = \frac{\sum_{\boldsymbol{i} \in A}(I_m(\boldsymbol{i}) - b(\boldsymbol{i}))}{F_{\hat{s}}}, \quad (11)$$

where $F_{\hat{s}}$ is the total flux of tPSF along the trajectory $\hat{s}(t)$.

In step 4, we use discrete numerical integration to construct the cost function $L$ and use an accelerated method to minimize $L$. More details can be found in the Appendix. The subsequent steps in the algorithm have not been detailed. The source code of our algorithm is available online at https://github.com/astrometry-jnu/TrailedSourceCentering and archived by Zenodo[5] (L.-P. Wu et al. 2024).

## 3. Experiment

This section evaluates the performance of our centroiding algorithm on trails generated from two representative trajectory types: circular arcs and irregular curves. These trails were created by convolving predefined trajectories with a PSF. While these trajectories do not encompass the full spectrum of motion patterns observed in real-world scenarios, they provide a valuable basis for assessing the algorithm's effectiveness. Circular arc trajectories are used to systematically investigate the algorithm's performance under varying conditions, such as changes in arc length, curvature, and FWHM. Irregular curves, on the other hand, are used to test the algorithm's robustness and adaptability in more complex scenarios.

### 3.1. Arc Trails

We generate a series of circular arcs as predefined trajectories. The length and central angle of each generated arc are adjusted within specific ranges: arc lengths vary from 20 pixels to 200 pixels in increments of 4 pixels, and central angles range from 0° to 180° in increments of 4°. Since the curvature at any point on a circular arc is equal to the central angle divided by arc length, a trajectory with a fixed length exhibits higher curvature as its central angle increases. Additionally, each trajectory is rotated and shifted by a subpixel bias to introduce randomness. Gaussian PSFs are then used to synthesize various trailing images. For each trajectory, different FWHM values are applied: 1.0, 1.5, 2.0, 2.5, and 3.0 pixels, respectively.

---
[5] doi:10.5281/zenodo.12679112.

Since our algorithm's performance is dependent on the initial control points, we constrained our analysis to the simplest case of three initial control points (the effect of additional control points is discussed later). Under this constraint, we investigated the algorithm's performance with respect to three parameters: central angle, trajectory length, and FWHM.

In the experiments, three initial control points were predefined for each trajectory, corresponding to the start point, midpoint, and endpoint of the circular arc. To better reflect real-world conditions, we introduced random perturbations of up to 3 pixels around these points. Subsequently, our centroiding algorithm was applied to determine the centroid of each trajectory. The accuracy of the centroiding algorithm was evaluated by calculating the Euclidean distance between the estimated centroids and the corresponding ground-truth centroids.

Our experimental results demonstrate a trend: for a fixed central angle, the centroiding accuracy tends to decrease as the arc length increases. At a critical arc length ($L_0$), the centroid determination accuracy abruptly deteriorates, preventing our algorithm from accurately locating the trail's centroid. Furthermore, this threshold $L_0$ exhibits a positive correlation with increasing FWHM. As illustrated in Figure 3, for a central angle of 120°, we observed $L_0$ values of approximately 64, 76, 92, 104, and 116 pixels, corresponding to FWHMs of 1.0, 1.5, 2.0, 2.5, and 3.0 pixels, respectively. For arc lengths below $L_0$, the centroiding accuracy fluctuates and decreases with increasing arc length, with a maximum deviation of about 0.1 pixels. However, beyond this threshold, the accuracy declines dramatically, resulting in an inability to determine the accurate centroid. This behavior can be attributed to the limitations of our initial three-point approximation of the trajectory. As the arc length increases, the initial trajectory diverges farther from the true trajectory. Consequently, the optimization process, which refines the trajectory through iterative steps, struggles to converge to the global optimum (i.e., the true trajectory). Instead, it may converge to a local optimum or fail to converge altogether, leading to inaccurate centroid estimates. Theoretically, increasing the number of control points for a more accurate initial trajectory could improve the convergence properties of our centroiding algorithm. However, this is challenging in automated processing.

As shown in Figure 3, the threshold $L_0$ increases with the increasing FWHM. This indicates that the algorithm's ability to converge to the global optimum using three initial control points improves as the FWHM increases. The increased thickness of the trail due to larger FWHM values provides more data points for the optimization process to refine the initial trajectory. This enhanced data availability facilitates a more accurate approximation of the true trajectory, thereby improving the algorithm's robustness to bigger FWHMs.

For a fixed arc length, our experiment shows similar trends: the accuracy of the centroid calculation exhibited a fluctuating decrease as the central angle increased. When the central angle reached a certain threshold $A_0$, the centroiding accuracy dropped sharply. This threshold $A_0$ also increased with increasing FWHM. Figure 4 illustrates the variation of the centroiding accuracy with central angles for different FWHM values when the arc length is fixed at 100 pixels.

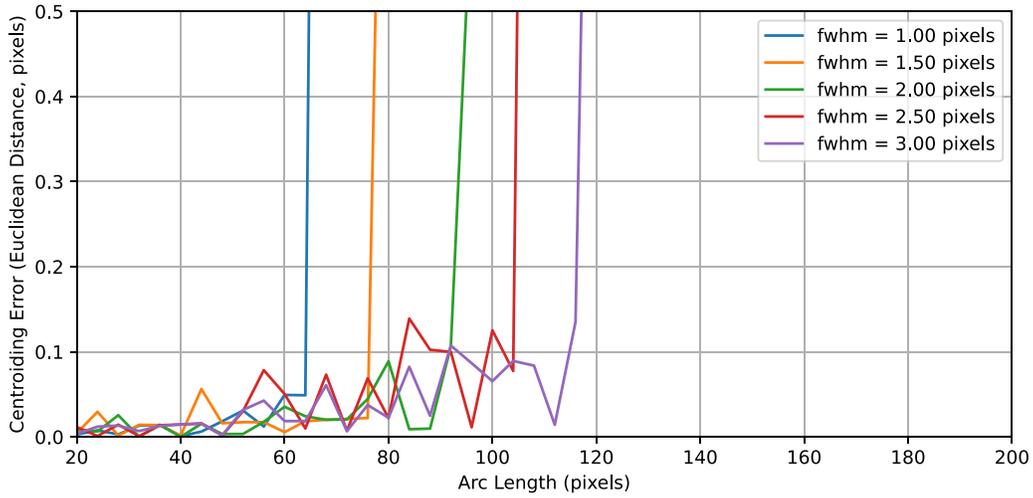

**Figure 3.** Centroiding error vs. arc length and FWHM at a fixed central angle of 120°. Here, the error is the Euclidean distance between the estimated and ground-truth centroids.

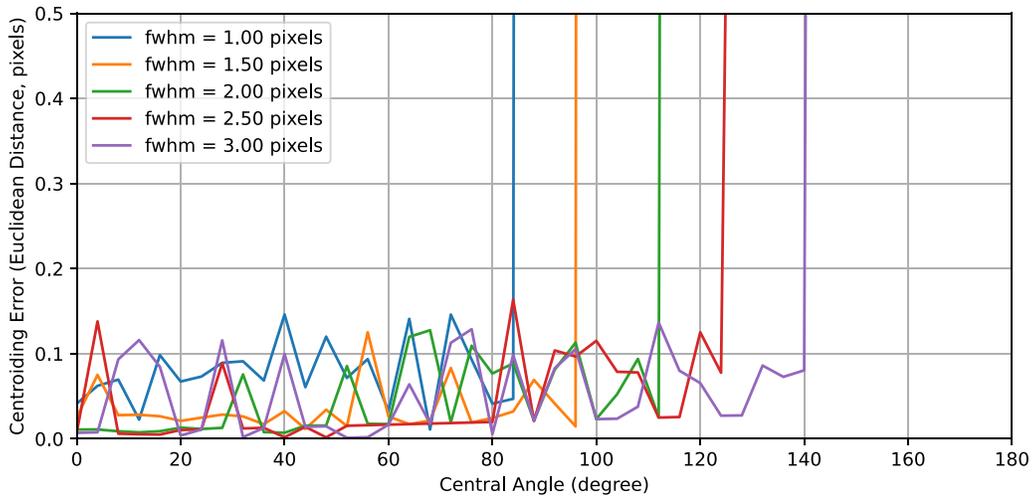

**Figure 4.** Centroiding error vs. central angle and FWHM at a fixed arc length of 100 pixels. The error is the Euclidean distance between the estimated and ground-truth centroids.

To further visualize the variations in centroiding accuracy, heat maps were generated showing the dependence of the centroiding accuracy on both the central angle and arc length for different FWHM values. Figure 5 shows them. These heat maps demonstrate the complex relationship between the accuracy of our centroiding algorithm and the arc length, central angle, and FWHM. In simpler terms, for smaller arc lengths and central angles, our algorithm converges to an accurate result more consistently, starting from the initial three control points. In most cases, it can achieve an accuracy of less than 0.2 pixels. Regardless of the FWHM, the white regions in the upper right corners of panels ((a)–(c)) indicate that for larger central angles and arc lengths, starting from three initial control points, the centroiding accuracy may exceed 1.0 pixel. Conversely, this suggests that in practical applications with larger central angles and arc lengths, increasing the density of the initial control points is recommended.

This simulation did not investigate the effects of the number of initial control points or their positions. There were two reasons for this: first, the number of possible variations in control points is vast, making it impractical to simulate and test all scenarios. Second, from a theoretical perspective, it is known that more accurate initial control points, placed closer to the true trajectory, facilitate a more efficient convergence to the optimal solution during the optimization process. This suggests that increasing the number of initial control points and placing them closer to the true trajectory can improve the accuracy of the results. While this is straightforward in manual operations, it poses challenges in automated processes. The study of three initial control points in this work provides a foundation for future automated processing applications.

### 3.2. Irregular Trails

In this section, we use simulated irregular trails to evaluate the performance of our centroiding algorithm in more complex scenarios. To generate irregular trails, we constructed 80 synthetic trajectories with an average length of ~40 pixels, based on observations from the Cassini ISS images. These trajectories exhibit diverse motion patterns, including linear but nonuniform motion, uniform motion along irregular curves, and irregular motion with varying speed. To introduce variability, we scaled and adjusted the speeds of these trajectories. Figure 6(a) presents three representative examples of these trajectories. The orange crosses mark the centroids of the trajectories, corresponding to the positions of the point sources at the mid-exposure moment. Note that these centroids may not coincide with the geometric centers of the curves due to nonuniform motion.

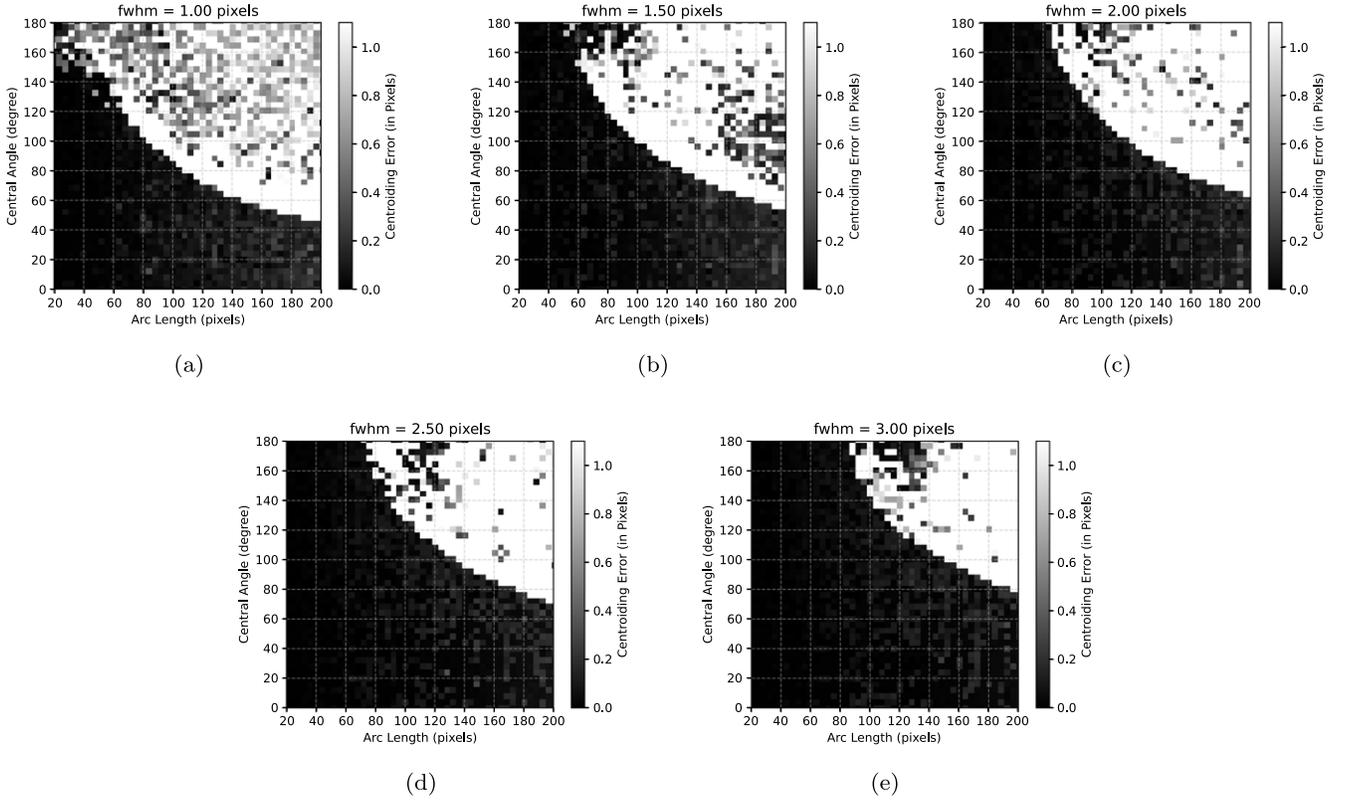

**Figure 5.** Variations in centroiding errors with both central angle and arc length for different FWHMs. (a) FWHM = 1.0 pixels. (b) FWHM = 1.5 pixels. (c) FWHM = 2.0 pixels. (d) FWHM = 2.5 pixels. (e) FWHM = 3.0 pixels.

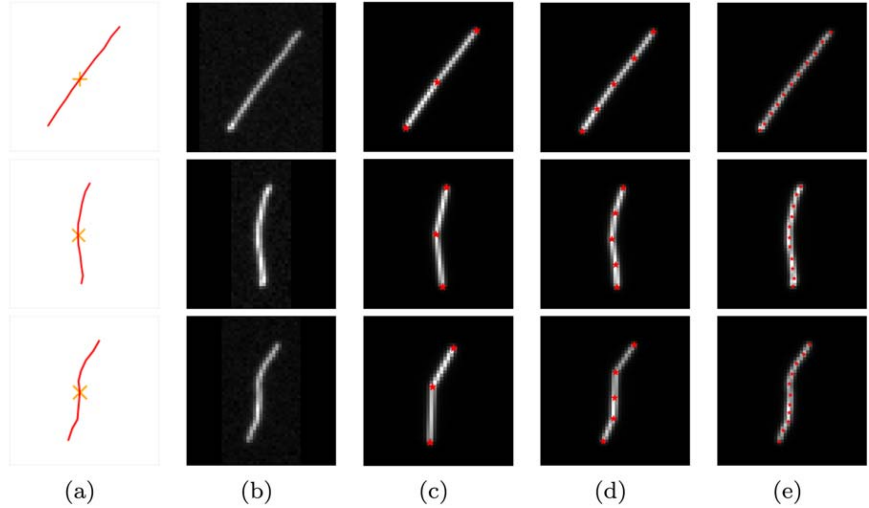

**Figure 6.** Simulated trails and fitting processes. (a) Predefined object trajectories: linear but nonuniformly moving (top); uniformly moving but irregularly curved (middle); and irregularly curved and nonuniformly moving (bottom). The cross marks the object's position at the midpoint of the exposure, $s(0)$. (b) Simulated trails (PSF convolution). (c) Initial control points and fitted trajectories. (d) Refined control points after one iteration. (e) Final control points and fitted trajectories.

After giving the motion trajectories and velocities of the point sources, we used the PSF of the Cassini ISS to construct the trailing images based on these trajectories. Since we are concerned with astrometry issues in ISS images, we specifically employed the ISS PSF rather than other PSFs. Additionally, we set the parameter $f_\star = 100$ and synthesized the foreground images following Equation (4). Figure 6(b) presents three generated trails, which correspond to the three trajectories shown in Figure 6(a).

To investigate the impact of noise on our algorithm's performance, we added different levels of noise to the 80

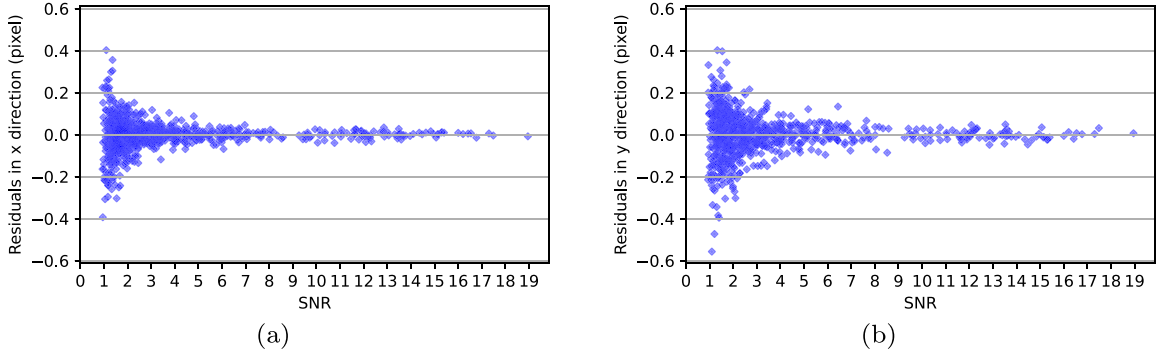

**Figure 7.** Residuals vs. SNR from the simulation experiments. (a) In the *x*-direction. (b) In the *y*-direction.

noiseless images, generating noisy images with varying SNRs. In general, the background of a CCD image is composed of bias, read noise, and dark current. The bias is simply a constant value, and the read noise is subject to Gaussian distribution, while the dark current varies among different circumstances. For simplicity, we treat all these components collectively as a Gaussian noise model, with a mean value of 100 and a standard deviation (SD) $\sigma_\eta = r_\eta \cdot f_\star$, where $r_\eta$ ranges from 0.05 to 0.5.

Finally, for the 80 trajectories, we created 800 images of trails. That is, we generated 10 images for each trajectory with different levels of noise. The noise levels were added at 5%, 10%, 15%, ..., 50% of $f_\star$. For these images, we also calculated their SNRs, which we define as follows:

$$\text{SNR} = \frac{\mu_s - \mu_b}{\sigma_b}, \quad (12)$$

where $\mu_s$ and $\mu_b$ denote the mean intensity of the trail and background, respectively, and $\sigma_b$ represents the SD of the background intensity. The SNRs of the noisy images ranged from about 1.0 to 19.0, ensuring a comprehensive representation of noisy image scenarios with varying levels of noise.

For the entire set of 800 images, our proposed centroiding algorithm was applied to measure the centroids of all these trajectories. We initialized the control points, which were then progressively refined using a coarse-to-fine approach. This iterative process yielded an optimized sequence of control points, enabling accurate determination of the trajectories and their centroids. Figures 6(c)–(e) illustrate the refined control points obtained through this process, corresponding to the original images shown in Figure 6(b).

After obtaining the centroids of the trajectories, we calculated the residuals between the computed centroids and the true ones. Figure 7 presents the distributions of these residuals under different noise levels. Figures 7(a) and (b) are the residuals in the *x*-direction and *y*-direction, respectively. It is observed that as the SNR increases, the accuracy of the results improves significantly in the *x*- and *y*-directions. Table 1 displays statistical results of the measurements across all 800 images. At high SNRs (>10), our centroiding algorithm achieves a high precision of 0.02 pixels in both directions, and the mean and SD of the distance error are approximately 0.03 pixels and 0.01 pixels, respectively. Even in low-SNR (1.0 < SNR < 1.1) conditions, the algorithm's precision reaches about 0.19 pixels. The mean values of these residuals in both directions are always close to zero, indicating the absence of any systematic error. In terms of the distance error, its mean is 0.23 pixels and SD is 0.12 pixels. Table 1 shows the

**Table 1**
Statistical Results of Centroiding Errors for Synthetic Irregular Trails with Varying SNRs

| SNR | $\Delta x$ (pixels) | | $\Delta y$ (pixels) | | $\Delta s$ (pixels) | |
|---|---|---|---|---|---|---|
| | Mean | SD | Mean | SD | Mean | SD |
| 1.0 ∼ 1.1 | −0.02 | 0.18 | −0.02 | 0.19 | 0.23 | 0.12 |
| 1.1 ∼ 1.3 | −0.00 | 0.14 | −0.01 | 0.15 | 0.17 | 0.11 |
| 1.3 ∼ 1.6 | −0.01 | 0.10 | 0.01 | 0.14 | 0.14 | 0.09 |
| 1.6 ∼ 2.0 | −0.00 | 0.08 | 0.00 | 0.12 | 0.11 | 0.07 |
| 2.0 ∼ 2.5 | −0.01 | 0.06 | −0.01 | 0.09 | 0.09 | 0.06 |
| 2.5 ∼ 3.0 | 0.01 | 0.04 | 0.00 | 0.07 | 0.07 | 0.04 |
| 3.0 ∼ 4.0 | 0.00 | 0.04 | −0.01 | 0.06 | 0.06 | 0.04 |
| 4.0 ∼ 5.0 | 0.00 | 0.03 | −0.00 | 0.04 | 0.06 | 0.03 |
| 5.0 ∼ 7.0 | −0.00 | 0.02 | 0.01 | 0.04 | 0.05 | 0.03 |
| 7.0 ∼ 10.0 | −0.00 | 0.02 | −0.00 | 0.03 | 0.04 | 0.02 |
| 10.0 ∼ 13.0 | −0.00 | 0.02 | −0.00 | 0.02 | 0.03 | 0.01 |
| >13.0 | 0.00 | 0.01 | −0.00 | 0.02 | 0.02 | 0.01 |

**Note.** $\Delta s$ is the distance error of the estimated centroid $\Delta s = \sqrt{\Delta x^2 + \Delta y^2}$.

precision in the *x*-direction is slightly better than that in the *y*-direction. It is probably because most of the synthetic images are trailed more severely in the *y*-direction than in the *x*-direction, as shown in Figure 6.

Furthermore, our algorithm not only determines the centroids of trails but also computes the relative motion trajectories of observed objects. We conduct a comparison between the estimated trajectories and their predefined ones to evaluate the algorithm's performance. The observation interval was defined as $[−T, T]$, where sampling points were established at $0.1T$ intervals, resulting in 21 temporal sampling points per trajectory. At each sampling point, we compared the estimated position against the ground truth. Subsequently, the mean errors and SDs were calculated across all sampling points and trajectories at different SNR levels. Table 2 presents the estimation errors for *x* and *y* coordinates separately, as well as the distance errors. The results reveal that at lower SNR levels (1.0 ∼ 1.1), the average distance error of the estimated trajectories reaches 0.17 pixels with an SD of 0.11 pixels. At higher SNR levels (10.0 ∼ 13.0), the average error and SD decrease significantly to 0.02 pixels and 0.01 pixels, respectively. The trajectory estimation accuracy demonstrates a clear positive correlation with increasing SNR values. Figure 8 illustrates the distributions of the trajectory estimation errors in both the *x*- and *y*-directions across the exposure span for all

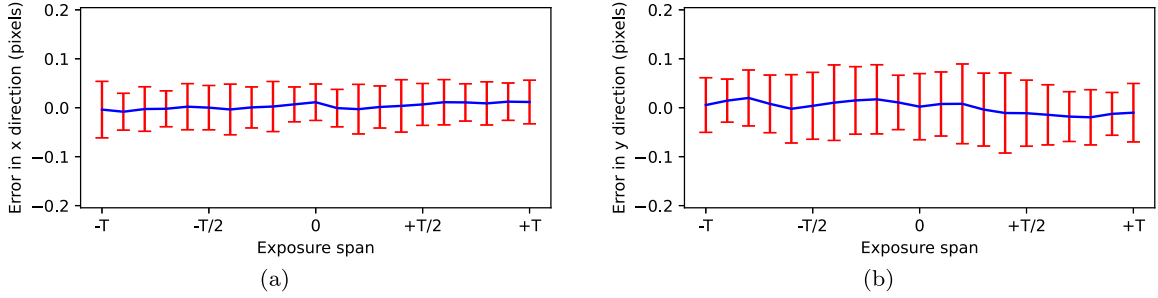

**Figure 8.** Trajectory estimation residuals at different exposure times under a noise level of $2.5 < \mathrm{SNR} < 3.0$. (a) Residuals in the *x*-direction. (b) Residuals in the *y*-direction. The horizontal axis represents the exposure span of the trajectories. The blue curves represent the mean residual for a specific time *t*. The red bars are $1\sigma$ error bars.

**Table 2**
Statistical Results of Trajectory Estimation Errors on Synthetic Trails with Varying SNRs

| SNR | $\Delta x$ (pixels) | | $\Delta y$ (pixels) | | $\Delta s$ (pixels) | |
|---|---|---|---|---|---|---|
| | Mean | SD | Mean | SD | Mean | SD |
| $1.0 \sim 1.1$ | 0.02 | 0.14 | −0.02 | 0.15 | 0.17 | 0.11 |
| $1.1 \sim 1.3$ | −0.01 | 0.12 | 0.00 | 0.14 | 0.15 | 0.11 |
| $1.3 \sim 1.6$ | −0.01 | 0.09 | 0.01 | 0.12 | 0.12 | 0.08 |
| $1.6 \sim 2.0$ | 0.00 | 0.07 | −0.01 | 0.10 | 0.10 | 0.07 |
| $2.0 \sim 2.5$ | 0.00 | 0.06 | 0.00 | 0.08 | 0.08 | 0.06 |
| $2.5 \sim 3.0$ | 0.00 | 0.04 | 0.00 | 0.07 | 0.07 | 0.04 |
| $3.0 \sim 4.0$ | 0.00 | 0.04 | 0.00 | 0.05 | 0.05 | 0.04 |
| $4.0 \sim 5.0$ | 0.00 | 0.03 | 0.00 | 0.04 | 0.04 | 0.03 |
| $5.0 \sim 7.0$ | 0.00 | 0.02 | 0.00 | 0.03 | 0.03 | 0.02 |
| $7.0 \sim 10.0$ | 0.00 | 0.02 | 0.00 | 0.03 | 0.03 | 0.02 |
| $10.0 \sim 13.0$ | 0.00 | 0.02 | 0.00 | 0.02 | 0.02 | 0.01 |
| $>13.0$ | 0.00 | 0.01 | 0.00 | 0.02 | 0.02 | 0.01 |

**Note.** $\Delta s$ is the distance error of the estimated trajectory $\Delta s = \sqrt{\Delta x^2 + \Delta y^2}$.

trajectories with noise levels of $2.5 < \mathrm{SNR} < 3.0$. The plots span the exposure interval $[-T, T]$, with panel (a) displaying the *x*-direction errors and panel (b) showing the *y*-direction errors. The blue curves represent the mean errors at each sampling point, while the red bars indicate $\pm\sigma$ error margins. The approximately symmetrical distribution of errors around zero suggests that the algorithm is unbiased, while the uniform error-bar lengths indicate stable precision throughout the exposure span. While these experiments cannot exhaustively cover all possible scenarios, the results provide compelling evidence for the algorithm's stability and effectiveness in trajectory estimation, consistently achieving subpixel accuracy within 0.1 pixels.

## 4. Application

The Cassini ISS consists of two fixed-focal-length cameras—the Narrow Angle Camera (NAC) and the Wide Angle Camera (WAC)—along with an array of filter pairs (C. C. Porco et al. 2004). The NAC was an $f/10.5$ reflecting telescope with an image scale of $\sim 6~\mu\mathrm{rad\,pixel}^{-1}$, a $0\overset{\circ}{.}35 \times 0\overset{\circ}{.}35$ field of view, and a spectral range from 200 nm to 1100 nm. It was equipped with 24 spectral filters, arranged in two wheels of 12 filters each. The WAC, designed as an $f/3.5$ refractor with a $3\overset{\circ}{.}5 \times 3\overset{\circ}{.}5$ field of view, was similarly equipped with a set of spectral filters. The PSFs of the ISS cameras depend on the combination of both the camera and the filter, which have been published by R. West et al. (2010) and B. Knowles et al. (2020). The ISS cameras can be operated in different binning modes: $1 \times 1$, $2 \times 2$, or $4 \times 4$.

During the Cassini mission, the ISS captured numerous images, which were used for the astrometry of planetary satellites (N. J. Cooper et al. 2006, 2014; R. Tajeddine et al. 2013, 2015; Q. F. Zhang et al. 2018, 2021, 2022). Some of these images exhibit trailed reference stars, due to extended exposure times while tracking the target. Figure 9 shows two examples of such images, in which the reference stars exhibit significant trailing, while the satellites appear as resolved disks. Traditionally, images with trailed reference stars have been discarded in astrometric analysis, because accurately determining the centroids of these stars is difficult. More specifically, it is challenging to accurately determine the positions of these trailed stars at the mid-exposure moment, which prevents precise camera-pointing correction. Without this correction, reliable measurement of the astrometric positions of the satellites becomes impossible. As a result, such images were typically excluded from analysis. The algorithm proposed in this paper addresses the challenges of traditional methods. It accurately determines the centroids of trailed reference stars, enabling precise camera-pointing correction. Subsequently, we apply a limb-fitting technique to determine the centroids of the disk-resolved satellites. By combining the corrected pointing with the satellite centroids, we can derive precise astrometric positions for the satellites. For this study, we selected a subset of Cassini ISS images captured by the NAC using (CL1, CL2) filters. These images, with exposure times ranging from 4.6 to 100 s, all exhibit trailed reference stars. Two representative examples from this data set are displayed in Figure 9. We applied our algorithm to these images and successfully determined the astrometric positions of the satellites. The following sections describe the process and results of this work.

To analyze these images, preprocessing is required to determine the centroid of each reference star in each image. Initially, a mask of the target satellite was generated using a false-stars-removal algorithm (Q.-F. Zhang et al. 2020), and the pixel intensities of the target satellite were set to zero using the mask. Second, in this image, a bright, high-SNR trailed star was selected, and our centroiding algorithm was applied to measure it. This measurement yielded both the center position of this star and a relative tPSF with respect to its center, which we used as a template. Finally, we employed this tPSF template through template matching and local peak detection to detect all the trailed stars in the image and determine their centroids in pixels. Please note that the previous step of setting the target region to zero reduced interference during the template-matching process. Ultimately, the trajectory was fitted to the detected trail by using the least-squares method to determine the centroids in subpixels.

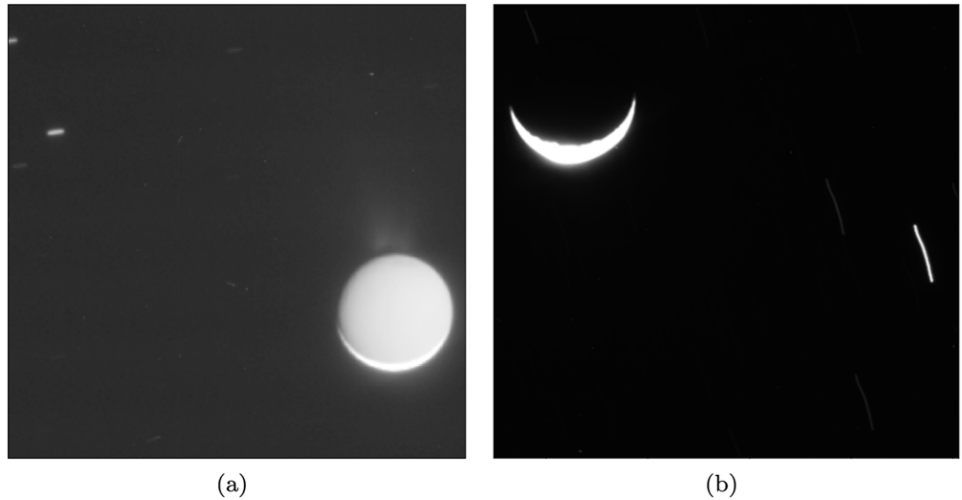

**Figure 9.** Subimages from Cassini ISS images. (a) Subimage of Enceladus, processed using a logarithmic transformation to enhance object visibility, where the trails of stars are short and uniform. (b) Subimage of Dione, where the trails of stars are long and irregular.

After preprocessing, we employed Caviar (N. Cooper et al. 2018), a software package dedicated to the astrometry of ISS images, to perform measurements on each image. First, the centroids of all the trailed stars in the image were loaded into Caviar and were taken as the reference stars' image positions. Then, based on the nominal pointing of the ISS image, the positions in the International Celestial Reference System (ICRS) of all possible stars within the field of view were retrieved from GAIA DR3 (Gaia Collaboration et al. 2023). All these catalog stars' positions were reduced into images to obtain their image coordinates. Subsequently, the catalog stars were matched with the image stars to correct the camera pointing at the midpoint of the exposure. After pointing correction, the limb-fitting method (R. Tajeddine et al. 2013) was used to measure the centers of the target satellites, because they are resolved disks in these images. Here, the edge-detection parameters were adjusted to ensure the detected contour encompasses the faint limb as much as possible, improving the accuracy of the limb fitting. The satellite center positions were corrected with the geometric distortion model given in W. Owen (2003). Finally, the astrometric positions of the satellites were reduced.

Finally, 267 images are reduced, including 188 images of Enceladus, 55 of Dione, 10 of Mimas, six of Rhea, and eight of Tethys. Table 3 gives a sample of all astrometric results from 267 observations. The full table is published in its entirety in machine-readable format.

## 5. Discussion

To assess the accuracy of the measurements, the astrometric positions of the targets in 267 ISS images were evaluated against the calculated positions from the JPL ephemeris SAT427, resulting in the $(O - C)$ residual for each observation. Figure 10 shows the residuals of all satellites relative to the JPL ephemeris SAT427 in the sample and line directions, in pixels, and in the $\alpha \times \cos\delta$ and $\delta$ directions, in kilometers.

Table 4 gives the mean and SDs of these residuals. From the table, we can find that the mean values are all close to zero. The SDs of the residuals for all targets are 0.19 and 0.18 pixels in the sample and line directions, respectively. In angular terms, the SDs in $\alpha \times \cos\delta$ and $\delta$ are $0.''23$. When translated to linear distances, the SDs in $\alpha \times \cos\delta$ and $\delta$ correspond to 1.33 km and 1.36 km, respectively. Our centroiding algorithm for trailed stars achieves positional precision comparable to those of existing methods for untrailed stars shown in Q. F. Zhang et al. (2021). These uncertainties primarily arise from two sources: measurement errors of trailing reference stars and satellite center determination. Approximately 2200 trailing reference stars were measured using our centroiding algorithm, achieving a precision of approximately 0.1 pixels. For disk-resolved satellite images, we employed a limb-fitting technique to determine the centers. This limb-fitting process introduces additional uncertainties. When combined with the reference star measurements, this results in a total positional uncertainty of approximately 0.2 pixels, equivalent to $0.''23$ in angular terms. It is important to note that the Cassini ISS NAC is an undersampled optical system, with an FWHM of approximately 1.3 pixels under the (CL1, CL2) filters, which inherently limits the achievable astrometric precision. This instrumental characteristic represents a fundamental constraint on the overall measurement accuracy of the data we processed.

In this study, 158 of all images were also analyzed in Q.-F. Zhang et al. (2023), using a streak-centering technique for linear trails. We compare the results of the two methods for these images. Table 5 presents a comparison of the results from both measurement techniques. The accuracy achieved by the method proposed in this paper is comparable to that of the previous method for linear trails. This finding suggests that the new method can effectively process linear trails. Additionally, this method can also measure irregular trails. When combined with the findings from Table 4, it is evident that the method maintains equivalent accuracy for both linear and irregular trails

**Table 3**
A Sample of Astrometric Results of 267 ISS Images of Several Saturnian Satellites

| Image ID | Midtime (UTC) | $\alpha_c$ (deg) | $\delta_c$ (deg) | Twist (deg) | Sample (px) | Line (px) | $\alpha$ (deg) | $\delta$ (deg) | Body |
|---|---|---|---|---|---|---|---|---|---|
| N1516306300_1 | 2006/Jan/18 19:42:03.747 | 340.0974427 | −3.2634021 | 84.4729103 | 563.996 | 594.149 | 340.0708770 | −3.2427305 | TETHYS |
| N1516373834_1 | 2006/Jan/19 14:27:36.382 | 319.2640714 | −1.2325407 | 173.7097725 | 566.929 | 574.516 | 319.2427618 | −1.2519572 | RHEA |
| N1665830854_1 | 2010/Oct/15 10:00:37.748 | 347.2422982 | 0.9019447 | 344.6098306 | 149.860 | 120.864 | 347.1475930 | 0.8319550 | DIONE |
| N1828211488_1 | 2015/Dec/07 19:27:34.403 | 76.0790129 | −4.3630243 | 2.9672921 | 509.983 | 893.152 | 76.0716908 | −4.2322167 | ENCELADUS |
| N1875919335_1 | 2017/Jun/11 23:33:18.593 | 77.1148032 | 2.2325330 | 3.9672694 | 509.074 | 776.780 | 77.1076659 | 2.3233203 | ENCELADUS |

**Note.** Column (1): image name. Column (2): the date and exposure midtime of the image (UTC). Columns (3)–(5): the R.A., decl., and twist angle of the camera's pointing vector in the ICRS-centered Cassini. Columns (6)–(7): the measured position of the target. Columns (8)–(9): the R.A. and decl. in the ICRS-centered Cassini for the target. Column 10: target name. The origin of the sample and line coordinate system are shown to the top left of the image, with the line y increasing downward and the sample x rightward. The full table is published in machine-readable format. A sample is shown here for guidance regarding its form and content.

(This table is available in its entirety in machine-readable form in the online article.)

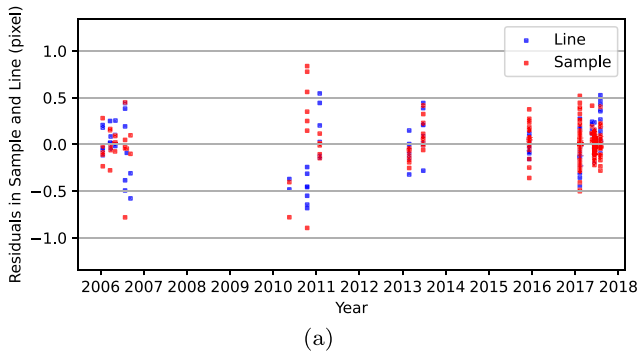
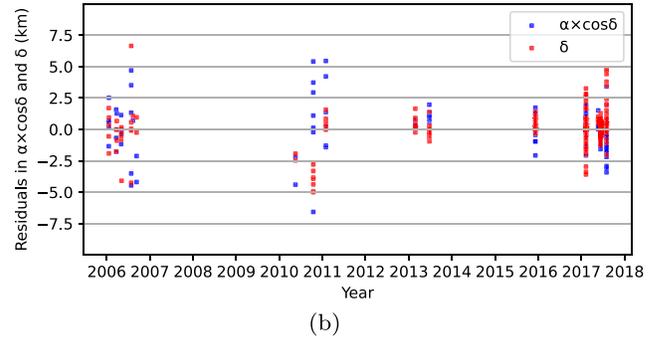

**Figure 10.** Residuals of our astrometry results for five satellites relative to the JPL ephemeris SAT427. (a) Residuals in the sample and line directions, in pixels. (b) Residuals in the $\alpha \times \cos\delta$ and $\delta$ directions, in kilometers.

**Table 4**
Statistics of All Astrometric Results

| Direction | Mean | SD |
| --- | --- | --- |
| Sample (pixels) | 0.01 | 0.19 |
| Line (pixels) | 0.00 | 0.18 |
| $\alpha \times \cos\delta$ (arcsec) | 0.02 | 0.23 |
| $\delta$ (arcsec) | 0.04 | 0.23 |
| $\alpha \times \cos\delta$ (km) | 0.06 | 1.33 |
| $\delta$ (km) | 0.18 | 1.36 |

**Note.** Mean values and SDs of the residuals of our results relative to the JPL ephemeris SAT427.

**Table 5**
Comparison with Previous Astrometric Results

| Direction | | Our Results (pixels) | Previous Results (pixels) |
| --- | --- | --- | --- |
| Sample | Mean | 0.022 | 0.003 |
|        | SD   | 0.192 | 0.193 |
| Line   | Mean | −0.009 | −0.036 |
|        | SD   | 0.155 | 0.156 |

**Note.** Mean values and SDs of the residuals from our astrometric results and those from the previous results by Q.-F. Zhang et al. (2023) for the same images.

## 6. Conclusions

This paper proposes a novel centroiding algorithm for point-source trails that can compute the centroids and reconstruct the trajectories of both linear and irregular trails generated by celestial objects moving uniformly or nonuniformly relative to the observing instrument. The algorithm first approximates a motion trajectory using a piecewise linear model with a set of control points as optimized parameters. A cost function is then constructed to optimize the set of control points, followed by a coarse-to-fine approach that iteratively refines the control points, allowing the piecewise linear model to progressively converge to the true trajectory. While the primary objective is centroid determination, this algorithm uniquely offers the capability of recovering the complete motion trajectory of the point source, a feature not present in existing centroiding algorithms. This distinct capability provides valuable insights into the relative motion between the observed object and the observing instrument.

We conducted simulation experiments on two types of trails: arc trails and irregular trails. The arc trail experiments demonstrated that the stability of our algorithm is correlated with the length, curvature, and FWHM of the trial. With three initial control points, the algorithm's stability decreases as the trial's length and central angle increase. Conversely, trails with larger FWHMs generally resulted in better algorithm stability. This finding suggests that for longer or more curved trails, it is recommended to increase the number and accuracy of the initial control points to improve the algorithm performance. The irregular trail experiments revealed that our algorithm achieves a centroiding precision of 0.02 pixels at a high SNR ($>10$) and 0.19 pixels at a low SNR ($1.0 \sim 1.1$).

While the primary objective of our algorithm is to compute the centroid of a trail, it also possesses the capability to reconstruct the entire motion trajectory of observed objects. Using the irregular trail experiments, we evaluated the algorithm's performance in reconstructing complete trajectories. For high-SNR conditions ($>10$), the average error in trajectory reconstruction is nearly 0 pixels, with an SD of 0.01 pixels. For low-SNR conditions ($1.0 \sim 1.1$), the average error increases to 0.11 pixels, with an SD of 0.07 pixels. These results indicate that our algorithm can effectively reconstruct entire trajectories, and the reconstruction error remains consistent over the entire observation period and across both the $x$ and $y$ dimensions.

The method was applied to the astrometric reduction of 267 real Cassini ISS images of Saturn's main inner satellites with trailed reference stars. The results were compared to JPL ephemeris SAT427 to get their residuals. The residuals of our measurement results have a mean and SD of 0.01 and 0.19 pixels in the sample direction, respectively, while in the line direction, they are 0.00 and 0.18 pixels, respectively. In terms of distance, our measurement precision is about 1.3 km. In particular, R. S. Park et al. (2024) point out that the most up-to-date ephemeris, SAT441, has a possible 1 km uncertainty. Hence, our actual measurement precision should be better than 1.3 km. Compared to the measurements obtained from images without trailed stars and those from images containing linear trailing, our method demonstrates comparable accuracy and precision. The results also indicate our method is applicable to irregularly curved trails.

The primary limitation of this algorithm is its computational cost, although we have accelerated the gradient computation. On an AMD R7 6800H 3.2 GHz CPU with 32GB of RAM, the trajectory computation time ranges from 3 s to 30 s per trajectory, depending on the trajectory's complexity. The individual trajectory processing time remains relatively long.

However, it is worth noting that in a single image containing multiple trailed sources, these objects often share the same tPSF, with only the intensity differing. Consequently, it is unnecessary to measure each object individually using the proposed centroiding algorithm. Instead, a tPSF template can be constructed from a well-imaged trailed source and used for template matching to identify the remaining trailed objects. This approach significantly reduces the processing time for an entire image. Another limitation of this algorithm is the requirement for the manual setting of the initial control points. While this manual intervention enhances the algorithm's stability when dealing with complex and irregular trails, it prevents fully automated processing. However, we foresee the potential for automated initialization, especially in specialized applications where tailored approaches can be developed.

In future work, we will incorporate trailed asteroid images from existing surveys (e.g., Pan-STARRS and the Zwicky Transient Facility) and Minor Planet Center submissions to assess the potential of our algorithm for asteroid studies. Additionally, we plan to explore extra tactics to accelerate the algorithm, improve its computational efficiency, and automate its execution. Furthermore, we intend to employ this algorithm for conducting astrometric reduction on ISS images of irregular Saturnian satellites. Observing irregular satellites from the Cassini ISS often requires long exposure durations, resulting in the reference stars and some irregular satellites appearing as trails in the images.

Overall, we propose a novel centroiding algorithm for trails. This algorithm can accurately determine the centroid of a trail and reconstructs its complete trajectory, regardless of whether the trail is linear or irregularly curved. Practical applications have validated the effectiveness of our algorithm.

## Acknowledgments

This work has been partly supported by the National Key R&D Program of China (No. 2022YFE0116800), the National Natural Science Foundation of China (No.12373073 and No. U2031104), and the Guangdong Basic and Applied Basic Research Foundation (No. 2023A1515011340 and No. 2024A1515011762). This work has made use of data from the European Space Agency (ESA) mission Gaia (https://www.cosmos.esa.int/gaia), processed by the Gaia Data Processing and Analysis Consortium (DPAC; https://www.cosmos.esa.int/web/gaia/dpac/consortium). Funding for the DPAC has been provided by national institutions, in particular the institutions participating in the Gaia Multilateral Agreement.

## Appendix
## Some Additional Implementation Details

To facilitate a deeper understanding of the implementation aspects of this algorithm, three additional details are explained here: constructing the cost function, accelerating the optimization, and handling the pixel binning.

### A.1. Constructing the Cost Function

The construction of $L(\hat{s})$ has three crucial terms: $\chi(i, \hat{s})$, $J_n(\hat{s})$, and $J_t(\hat{s})$. For the $\chi(i, \hat{s})$ term, we can use Equation (5) substituted into Equation (10) to obtain it. As for the two energy terms $J_n(\hat{s})$ and $J_t(\hat{s})$, according to Equation (8), we know they represent the integration of the second derivatives of $\hat{s}(t)$ in different directions. Since a piecewise linear model is employed, the nonzero second derivatives occur at the control points. Therefore, in the discrete form, Equation (8) can be written as

$$\begin{cases} J_n(\hat{s}) = \frac{1}{Q-1}\sum_{j=1}^{Q-1} \|\hat{s}''(t_j) \cdot e_n(t_j)\|^2 \\ J_t(\hat{s}) = \frac{1}{Q-1}\sum_{j=1}^{Q-1} \|\hat{s}''(t_j) \cdot e_t(t_j)\|^2 \end{cases}, \quad (A1)$$

where

$$\begin{cases} \hat{s}''(t_j) = \dfrac{x_{j-1} + x_{j+1} - 2x_j}{(2T/Q)^2} \\ e_t(t_j) = \dfrac{x_{j+1} - x_{j-1}}{\|x_{j+1} - x_{j-1}\|} \\ e_n(t_j) = (e_t^y(t_j), -e_t^x(t_j)) \end{cases}, \quad (A2)$$

where, $e_t^x(t_j)$ and $e_t^y(t_j)$ represent the components of $e_t(t_j)$ in the x- and y-directions in an image, respectively.

In Equation (7), there are two coefficients, $\lambda_n$, $\lambda_t$. They can be given by the user. In our case of analyzing ISS images, the empirical values of $\lambda_n$, $\lambda_t$ in Equation (7) are 0.09 and 0.01, respectively.

### A.2. Accelerating the Optimization

The main challenge of our method is the high computational cost of minimizing the cost function compared to classical centroiding algorithms. Gradient descent should be utilized, but if the gradients are approximated by finite differences, it might take several minutes to iterate and refine a trajectory (on an AMD R7 6800H 3.2 GHz CPU with 32GB of RAM). Fortunately, we can take advantage of PyTorch (A. Paszke et al. 2019), whose capability of autogradient computation works as an alternative to finite difference, which greatly reduces the complexity of the optimization. In this way, the processing time of a trajectory can be reduced to a few seconds. For details, see also our source code in GitHub: https://github.com/astrometry-jnu/TrailedSourceCentering.

Besides, various optimizers are tried. Different methods based on gradient descent have different times. It is found that RMSprop (T. Tieleman 2012), with a learning rate of 0.01, could work well, and we set the tolerance of a single control point's coordinate to 0.001 pixel.

### A.3. Handling the Pixel Binning

In practical scenarios, cameras sometimes use binning to increase light sensitivity. Binning combines neighboring pixels into a larger superpixel. For instance, Cassini ISS can use $2 \times 2$ or $4 \times 4$ binning. Binning changes the pixel coordinates. However, we tend to use a consistent coordinate system for all images. Let $i_\beta$ be the row–column index of a pixel after binning ($\beta$ denotes the binning factor, e.g., 2 for $2 \times 2$). The corresponding pixel's location in the unbinned frame is given by

$$i = \beta i_\beta + \frac{\beta - 1}{2}. \quad (A3)$$

This equation transforms any binned pixel coordinate to its equivalent in the unbinned frame.

The original PSF is no longer directly applicable to binned images, because we cannot perfectly recover the unbinned data. This means that the measured intensity of any pixel, $I_m(\boldsymbol{i})$, always reflects the post-binning state. Therefore, the cost function of Equation (7) must account for binning by using a post-binning PSF:

$$\psi_\beta(\boldsymbol{x}) = \sum_{\boldsymbol{d} \in \Omega} \psi(\boldsymbol{x} + \boldsymbol{d}), \quad (A4)$$

where $\boldsymbol{x}$ is an arbitrary coordinate (unbinned frame), and $\Omega$ is the set of displacement vectors from a large binned pixel's geometric center to those of small unbinned pixels. For example, the displacement vectors of $2 \times 2$ binning are $\Omega = \left\{ \begin{bmatrix} -0.5 \\ -0.5 \end{bmatrix}, \begin{bmatrix} -0.5 \\ 0.5 \end{bmatrix}, \begin{bmatrix} 0.5 \\ -0.5 \end{bmatrix}, \begin{bmatrix} 0.5 \\ 0.5 \end{bmatrix} \right\}$. Note that all coordinates and vectors mentioned above, except the row–column index $\boldsymbol{i}_\beta$, are defined in the unbinned frame.

### ORCID iDs

Linpeng Wu  https://orcid.org/0009-0009-8676-6931
Qingfeng Zhang  https://orcid.org/0000-0003-4086-9678
Valéry Lainey  https://orcid.org/0000-0003-1618-4281